\documentclass[twocolumn,showpacs,showkeys,preprintnumbers,amsmath,amssymb]{revtex4}
\usepackage[dvips]{graphicx}
\newcommand{\be}{\begin{equation}}
\newcommand{\ee}{\end{equation}}
\newcommand{\beq}{\begin{equation}}
\newcommand{\eeq}{\end{equation}}
\newcommand{\bea}{\begin{eqnarray}}
\newcommand{\eea}{\end{eqnarray}}
\newcommand{\eml}{\end{mathletters}}
\newcommand{\nn}{\nonumber\\}

\newcommand{\cm}[2]{\left[#1,#2\right]}

\newcommand{\avg}[1]{\left\langle  #1 \right\rangle}
\newcommand{\ket}[1]{|  #1 \rangle}
\newcommand{\bvac}{| 0)}
\newcommand{\fermi}{| {\text{HF}}\rangle}
\begin{document}

\title{Unified description of pairing and quarteting correlations\\ within the particle-hole-boson approach }

\author{V.V. Baran$^{1,2,3}$ and D.S. Delion$^{1,3,4}$}
\affiliation{$^1$"Horia Hulubei" National Institute of Physics and 
Nuclear Engineering, \\ 
30 Reactorului, RO-077125, Bucharest-M\u agurele, Rom\^ania \\
$^2$ Faculty of Physics, University of Bucharest,
405 Atomi\c stilor, POB MG-11, Bucharest-M\u agurele, RO-077125, Rom\^ania \\
$^3$Academy of Romanian Scientists,
54 Splaiul Independen\c tei, RO-050094, Bucharest, Rom\^ania \\
$^4$Bioterra University, 81 G\^arlei, RO-013724, Bucharest, Rom\^ania}

\begin{abstract}
We study the description of single-species and isovector pairing correlations  in the framework of the projected-BCS (PBCS) and the Quartet Condensation Model (QCM) from a particle-hole perspective and we introduce the representation of the QCM quartet condensate state in terms of particle-hole excitations with respect to the Hartree-Fock state. We also present a new bosonic approximation for both PBCS and QCM. 
In each case, the starting point is the reformulation of the pair/quartet condensate state in terms of 
particle-hole excitations with respect to the Hartree-Fock state. 
The main simplification of our approach is the assumption that the pair operators corresponding 
to both particle and hole states obey bosonic commutation relations. 
This simplifies tremendously the computations and allows for an analytic derivation of the averaged Hamiltonian
on the condenstate state as a function of the mixing amplitudes. We study both the pure bosonic approach 
and the renormalized version, and compare the particle-hole bosonic version to the naive prescription of applying the boson approximation directly to the original condensate state. We compare the fermionic and the renormalized particle-hole bosonic approach in the case of a picket fence model of doubly degenerate states and in a realistic shell model space with an effective interaction for the $N=Z$ nuclei above $^{100}$Sn.
\end{abstract}



\maketitle

\section{Introduction}

The $\alpha$-cluster model of the nucleus was proposed in order to 
explain the relative stability of $4n$ light nuclei \cite{Haf38,Ike68}.
The main theoretical difficulty is connected to strong antisymmetrisation
effects between nucleons entering $\alpha$-like structures.
Various microscopic $\alpha$-clustering models were proposed to account for it 
\cite{Flo63,Bri66,Ari71,Wil77,Fre97,Del02,Hor04,Fun09,Toh17}. 
At the same time, a simplified version considering proton and neutron pairs 
within the boson approximation was successful in explaining the even-odd pair 
staggering of binding energies \cite{Gam83}.
In medium and heavy nuclei the $\alpha$-clustering can experimentally be related 
to the $\alpha$-decay phenomenon \cite{Del10}.
It became clear that an $\alpha$-clustering component was necessary in addition to
the standard single-particle basis in order to describe the absolute value
of the $\alpha$-decay width \cite{Var92,Del04}. This can be explained
by the fact that $\alpha$-particles can appear only at relative low nuclear 
densities \cite{Rop98}, a situation which may be realised on the nuclear surface
of $\alpha$-decaying nuclei \cite{Del13}.

 This may be also the case for special configurations like the Hoyle state in $^{12}$C, which may be seen as a loosely bound agglomerate of three $\alpha$ particles condensed, as bosons, in the $0S$ orbit of their own cluster mean field. The understanding of the dynamics of $\alpha$ clusters in such situations was greately improved by the recent THSR approach \cite{Toh01}, which also triggered a significant amount of new interest in the field; for a recent review, see e.g. \cite{Sch16}. A bosonic type condensation is however not perfectly realized for clustered finite fermion systems, as generally there are significant residual  manifestations of the Pauli exclusion principle (e.g. the generic picture of a dilute gas-like $\alpha$-particle structure of the Hoyle state is accurate only to about 70\%-80\%, as evidenced by a variety of other treatments \cite{Mat04,Yam05,Laz11,Ish14}).  Nevertheless, an attractive feature of the THSR approach is that it exhibits the two opposite limits, namely that of a pure Slater determinant  and the other in which "the $\alpha$ particles are so far apart from one another that the Pauli principle can be neglected leading to a pure product state of $\alpha$ particles, i.e., a condensate"\cite{Sch16}.
 
Also recently, the simple Quartet Condensation Model (QCM) was proposed for the study of isovector pairing and 
quarteting correlations in $N=Z$ nuclei \cite{San12a,danielphd} and was further developed in 
Refs. \cite{San12,Neg14,San14,San15,Neg17,Neg18} to the 
case of isoscalar pairing and $N>Z$ nuclei. More general microscopic quartet models for a 
shell-model basis with an effective Hamiltonian were also recently developed \cite{Fu13,Fu15,Sam15,Sam15a,Sam16,Sam17}. In the quarteting type approaches, the basic building blocks are not the Cooper pairs anymore, but four-body structures composed of two neutrons and two protons coupled to the isospin $T = 0$ and to the angular momentum $J = 0$, denoted "$\alpha$-like quartets". The QCM approach was proven to be a very precise tool for the description of the amount of correlations present in the ground state of $N=Z$ nuclei. The antisymmetrization effects are significant in these configurations and thus the realization of an $\alpha$ condensation picture, in the sense mentioned above, is still an open question. Moreover, studies yet to be published \cite{San18} interestingly indicate the presence of "long-range correlations of condensate type" deduced from the behavior of the eigenvalues of the 4-body density matrix. 
 
Having said this, let us specify that in this paper we will use the terms "pair condensate" and "quartet condensate" to denote the specific projected-BCS (PBCS) and QCM trial states of Eqs. (\ref{PBCS0})  and (\ref{condensate}), also due to their structural similarity.
Anyway, we should keep in mind that an actual $\alpha$ condensate appears only at low densities, as opposed to the usual pair condensate.

 It is noteworthy that there are inherent difficulties in describing even the simpler pairing correlations, which have led to significant efforts dedicated to formulate approximate descriptions, including RPA \cite{Duk03} and coupled clusters methods \cite{Joh13,Hen14, Qiu19}. Recently, an improved approximate treatment of pairing correlations has been developed  \cite{Duk16}, in which the starting point is the reformulation of the PBCS condensate in the particle-hole basis. Particle-hole treatments have also been recently analyzed in the case of arbitrary generalized seniority cases \cite{Jia16}. 
 
 In the present work we take the opportunity of generalizing these ideas to the more complicated quartet correlations. We will argue that the particle-hole description is natural for both PBCS and QCM models, as can be seen from the behavior of the mixing amplitudes solutions (see Fig. \ref{fig1} below). We are thus motivated to find the representation of the QCM quartet condensate state in terms of particle-hole excitations with respect to the Hartree-Fock state (see Eqs. (\ref{phqcm1}-\ref{phqcm2}) below).

 We also introduce a new approximate hybrid fermionic-bosonic approach, which we will refer to as the \emph{particle-hole bosons} approach, applicable to both pairing and quarteting cases for the study of ground state correlations. In a first step, this approach requires the reformulation of the condensate state with respect to the correlated Hartree-Fock  vacuum, as opposed to the empty vacuum state $\ket{0}$. This ensures that a significant amount of fermionic correlations are already accounted for if we pass to bosonic degrees of freedom, but keep the same structure of the trial state. As it turns out, if we consider as a second step the simplest mapping of the individual pair operators to bosons, the ground state correlations in both pairing and quarteting cases are reproduced rather well as compared to the fully fermionic setting.  
 
 Although the basic ideas of treating quartet correlations in a boson formalism (see e.g. \cite{Gam83}, \cite{Dob98,Ler06,Nik17,Sam18}) and also considering the particle-hole excitations as bosons are certainly not new (see e.g. \cite{Kle91} for a thorough review on boson mappings), we are unaware of the two-step approach having been implemented in the specific way mentioned above.
 
 Let us finally note that in both pairing and quarteting cases our formalism is structurally very similar, leading to the same functional form of the energy of the bosonic condesate, up to form factors (see Eq. (\ref{universalH})  below). It is rather pleasing that in this sense a unified description of the pairing and the (significantly more complicated) quarteting correlations has been possible.
 
 Our work is structured as follows: in the following section we present the details regarding the reformulation of the pair and quartet condensates as particle-hole expansions. In section II B we develop the bosonic formalism, which is compared to the fully fermionic results in Section III. Finally, in Section IV we draw Conclusions.

\section{Theoretical background}

\subsection{Particle-hole representation of the pair and quartet condensates}

Let us consider first a model of a number $N_{lev}$ of doubly degenerate levels $i,\bar{i}$ (the so-called picket fence model), with  single particle energies $\epsilon_i$, where the ground state of the standard pairing Hamiltonian
\beq
\label{pairingH}
H=\sum_{i=1}^{N_\text{lev}} \epsilon_i \left(c^\dagger_i c_i+c^\dagger_{\bar{i}} c_{\bar{i}}\right)+\sum_{i,j=1}^{N_\text{lev}}V_{ij}P^\dagger_i P_j~,
\eeq 
is taken to be the PBCS pair condensate of $n_p$ pairs,
\beq
\label{PBCS0}
\ket{PBCS}=\left(\Gamma^{\dagger}(x)\right)^{n_p}\ket{0}~.
\eeq
Here, the coherent pair is a superposition of single particle pairs $P^{\dagger}_i=c^{\dagger}_ic^{\dagger}_{\bar{i}}$,
\beq
\Gamma^{\dagger}(x)=\sum_{i=1}^{N_\text{lev}} x_i P^{\dagger}_i~,
\eeq
and $\ket{0}$ is the vacuum state with no particles. We assume $N_{\text{lev}}>n_p$. All other notations are standard.

Following \cite{Rin80,Duk16}, instead of expressing the $\ket{PBCS}$ state with respect to the $\ket{0}$ vacuum, we may find an equivalent form 
involving the Hartree-Fock state
\beq
\label{fermi1}
\fermi=\left(\prod_{i=1}^{n_p}P^{\dagger}_i \right)\ket{0}~.
\eeq
To this end, we first decompose the coherent pair on components below and above the Fermi level as follows
\beq
\Gamma^\dagger(x)=\sum_{i=1}^{n_p} x_i P^{\dagger}_i+\sum_{i=n_p+1}^{N_\text{lev}} x_i P^{\dagger}_i \equiv \Gamma^\dagger_h(x)+\Gamma^\dagger_p(x)
\eeq
It is not difficult to show that the action of the hole component of the coherent pair of arguments $x$ on the $\ket{0}$ vacuum may be related to the action of the coherent pair of inverse arguments $1/x$ on the Hartree-Fock state (see Appendix A for computational details). In this way, one may prove that the reformulation of the pair condensate reads
\beq
\label{PBCS}
\begin{aligned}
\ket{PBCS}&=n_p! \cdot\Pi_1  \sum_{j=0}^{n_p} \frac{1}{(j!)^2}\left(\Gamma^\dagger_p(x)~\Gamma_h\left(\dfrac{1}{x}\right)\right)^j\fermi,
\end{aligned}
\eeq
where  $\Pi_1=x_1x_2\cdots x_{n_p}$.\\

This approach can be generalised from pair to quartet correlations.
To this purpose we consider the isovector pairing Hamiltonian applicable to both spherical and deformed nuclei
\beq
\label{isovectorH}
H=\sum_{i=1}^{N_\text{lev}}\epsilon_i\left(  N_{i, \pi}+ N_{i, \nu}\right)+\sum_{\tau=0,\pm1}\sum_{i,j=1}^{N_\text{lev}}V_{ij}P^{\dagger}_{i,\tau } P_{ j,\tau}~,
\eeq
where $\tau=0,\pm 1$ is the isospin projection. All other notations are identical to the pairing case. Within the QCM, one first defines a set of collective $\pi\pi$, $\nu\nu$ and $\pi\nu$ Cooper pairs
\beq
\Gamma^\dagger_\tau(x)\equiv\sum_{i=1}^{N_\text{lev}}x_i P^{\dagger}_{\tau, i}~,
\eeq
where the mixing amplitudes $x_i$ are the same in all cases due to isospin invariance. 
A collective quartet operator is then built by coupling two collective pairs to the total isospin $T=0$
\beq
Q^\dagger\equiv\left[\Gamma^\dagger\Gamma^\dagger\right]^{T=0}_{S=0}\equiv 2\Gamma^\dagger_1\Gamma^\dagger_{-1}-\big(
\Gamma^\dagger_0\big)^2~.
\eeq
Finally, the ground state of the Hamiltonian (\ref{isovectorH}) is described as a condensate of such $\alpha$-like quartets
\beq
\label{condensate}
| \Psi_{q}(x) \rangle=\big(Q^{\dagger}\big)^{q}|0\rangle~,
\eeq
where $q$ is the number of quartets. By construction, this state has a well defined particle number and isospin. Its structure is defined by the mixing amplitudes $x_i$, which are determined numerically by the minimization of the Hamiltonian expectation value, subject to the unit norm constraint.\\

In analogy with the standard pairing case described above, instead of expressing the quartet condensate state with respect to the $\ket{0}$ vacuum, we may find an equivalent form involving the Hartree-Fock state, in this case given by
\beq
\label{fermi2}
\fermi=\left(\prod_{i=1}^{q}P^{\dagger}_{1,i}P^{\dagger}_{-1,i} \right)\ket{0}~.
\eeq
The coherent pairs may be decomposed on components below and above the Fermi level
\beq
\Gamma^\dagger_\tau(x)=\sum_{i=1}^{q} x_i P^{\dagger}_{\tau,i}+\sum_{i=n_p+1}^{N_\text{lev}} x_i P^{\dagger}_{\tau, i} \equiv \Gamma^\dagger_{\tau, h}(x)+\Gamma^\dagger_{\tau, p}(x)~.
\eeq
As a consequence, the collective quartet decomposes as follows
\beq
 \label{quartetexpans}
 \begin{aligned}
&Q^\dagger(x)= 2\Gamma^\dagger_1\Gamma^\dagger_{-1}-\big(
\Gamma^\dagger_0\big)^2\\
&=2\Gamma^\dagger_{1,h}\Gamma^\dagger_{-1,h}-\big(
\Gamma^\dagger_{0,h}\big)^2+2\Gamma^\dagger_{1,p}\Gamma^\dagger_{-1,p}-\big(
\Gamma^\dagger_{0,p}\big)^2
\\&+2\left(\Gamma^\dagger_{1,p}\Gamma^\dagger_{-1, h}+\Gamma^\dagger_{-1,p}\Gamma^\dagger_{1, h}-\Gamma^\dagger_{0,p}\Gamma^\dagger_{0, h}\right)\\
&\equiv Q^\dagger_h(x)+Q^\dagger_p(x)+2\left[\Gamma^\dagger_{p}(x)\Gamma^\dagger_{ h}(x)\right]~.
\end{aligned}
\eeq
Given the more complicated decomposition of the quartet operator with respect to the simple pairing case, it is remarkable that the quartet condensate state may also be expressed as a particle-hole expansion.
The computational strategy is similar to the PBCS case, involving the introduction, for the hole subspace, 
of collective pair annihilation operators having as arguments the inverse amplitudes. The derivation is rather long and tedious, only its main points being presented in Appendix A. 
The exact, fully fermionic, analytical expression for the quartet condensate of Eq. (\ref{condensate}) as a particle-hole expansion reads
\beq
\label{phqcm1}
\begin{aligned}
\ket{\Psi_q}=&~2^q~ q!~\Pi_2\sum_{a=0}^q\sum_{b=0}^q
\lambda_{ab}\left(Q^\dagger_p(x) Q_h\left(\frac{1}{x}\right)\right)^a\\
&\times \left[\Gamma^\dagger_{p}(x)\Gamma_{ h}\left(\frac{1}{x}\right)\right]^b\fermi~,
\end{aligned}
\eeq
where
\beq
\label{phqcm2}
\begin{aligned}
\lambda_{ab}&=\frac{1}{2^a~b!}
\sum_{r=\text{Max}(0,N_{ab}-q)}^{a}\frac{(q-N_{ab})_{a-r}}{2^r (a-r)! (r!)^2}\\
&\times
\frac{\Gamma\left(\frac{3}{2}+q-r\right)}{\Gamma\left(\frac{3}{2}+N_{ab}-r\right)}~,
\end{aligned}
\eeq
and $\Pi_2=x_1^2x_2^2\cdots x_{q}^2$. The above formula is expressed using the total number of pair excitations in a given term $N_{ab}=2a+b$, the Gamma function $\Gamma(z)$ (not to be confused with the collective pair operator) and the Pochammer symbol $(z)_k=z(z-1)...(z-k)$. We also used a similar notation to that in Eq. (\ref{quartetexpans}) for the coupling of two pairs to $T=0$:
\beq
\left[\Gamma^\dagger_{p}(x)\Gamma_{ h}\left(\frac{1}{x}\right)\right]\equiv \sum_{\tau=\pm1, 0}\Gamma^\dagger_{\tau,p}(x)\Gamma_{\tau, h}\left(\frac{1}{x}\right)~.
\eeq
In the notation $\ket{\Psi_q}=\Pi_ 2\mathcal{O}_q \fermi$, some particular expressions for the operators $\mathcal{O}_q$ are
\begin{widetext}
\beq
\begin{aligned}
\mathcal{O}_1&=2 \left[\Gamma ^\dagger_p\Gamma
   _h\right]+\frac{1}{3}\left(Q^\dagger_pQ_h\right)+3\\
\mathcal{O}_2&=4 \left[\Gamma ^\dagger_p\Gamma _h\right]^2+20
   \left[\Gamma ^\dagger_p\Gamma
   _h\right]+\frac{1}{30
   }\left(Q^\dagger_pQ_h\right)^2+\frac{4 }{5}\left[\Gamma ^\dagger_p\Gamma _h\right]
   \left(Q^\dagger_pQ_h\right)+2
   \left(Q^\dagger_pQ_h\right)+30\\
\mathcal{O}_3&=8 \left[\Gamma ^\dagger_p\Gamma _h\right]^3+84
   \left[\Gamma ^\dagger_p\Gamma _h\right]^2+420
   \left[\Gamma ^\dagger_p\Gamma
   _h\right]+\frac{1}{63
   0}\left(Q^\dagger_pQ_h\right)^3+\frac{3 }{35}\left[\Gamma ^\dagger_p\Gamma _h\right]
   \left(Q^\dagger_pQ_h\right)^2\\
   &-\frac{99
   }{70}\left(Q^\dagger_pQ_h\right)^2+\frac{12
  }{7} \left[\Gamma ^\dagger_p\Gamma _h\right]^2
   \left(Q^\dagger_pQ_h\right)+12 \left[\Gamma
   ^\dagger_p\Gamma _h\right]
   \left(Q^\dagger_pQ_h\right)+114
   \left(Q^\dagger_pQ_h\right)+630~.
\end{aligned}
\eeq
The expressions for $\mathcal{O}_1$ and $\mathcal{O}_2$ have been checked by evaluating each individual term with the help of the symbolic computer algebra system Cadabra 2 \cite{cdb1,cdb2,cdb3}. It is very interesting to note that the quartet condensate actually arises as an interplay of isoscalar particle-hole excitations made out of coupled-pair type excitations and excitations of quartet-quartet type.\\\\

Notice that both Eqs. (\ref{PBCS}) and (\ref{phqcm1}) suggest that the pair mixing amplitudes corresponding to hole states behave inversely to those corresponding to particle states. 
Indeed, the mixing amplitudes corresponding to hole states and the inverse amplitudes 
corresponding to particle state both present an almost perfect linear behavior, as can be seen in Fig. \ref{fig1}, for the case of a constant pairing interaction strength. 
This was an early argument of the adequacy of a particle-hole description for both type of correlations. Moreover, the $x$ versus $1/x$ symmetry for particle and hole states is actually connected, in the PBCS case, to the mixing amplitudes $x$ being expressed in terms of the ratio between occupancy and vacancy BCS coefficients $u$ and $v$, in the context of the BCS state projection to good particle number. This fact opens the possibility to introduce a quasiparticle representation also for the case of quarteting correlations. \\

\begin{figure} [ht]
\begin{center}
\includegraphics[width=\columnwidth]{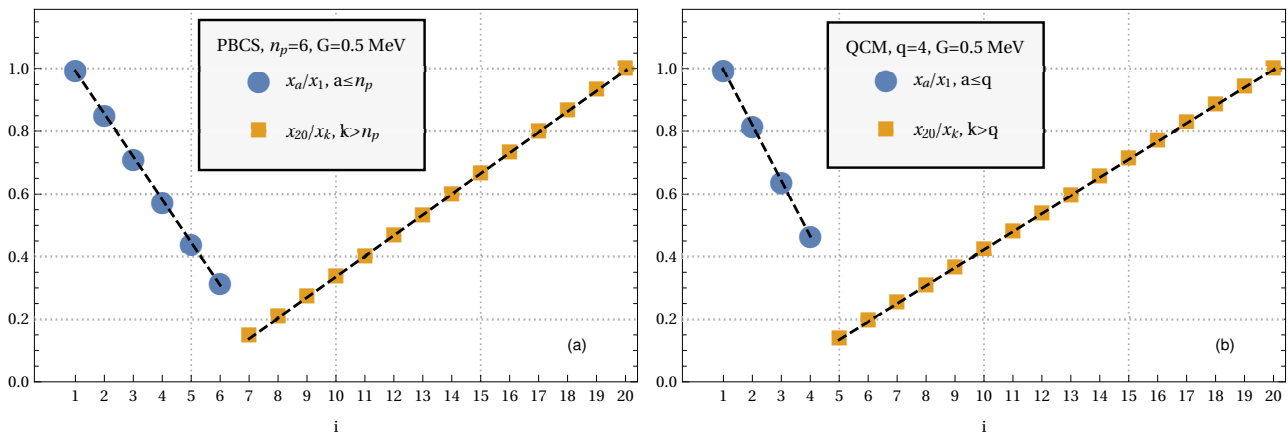}
\caption{The linear behavior of the normalized PBCS (a) and QCM (b) pair mixing amplitudes for hole states $x_{a}$ (circles) and \emph{inverses} 
of pair mixing amplitudes for particle states $1/x_{k}$ (squares) for a picket fence model of doubly degenerate levels $\epsilon_k=k$ MeV, 
$k=0,1,\dots,19$, for a coupling constant $G=0.5$MeV. The linear fit for each set of amplitudes is shown with dashed lines.}
\label{fig1}
\end{center}
\end{figure}

\end{widetext}

The expressions of Eq. (\ref{PBCS}) and those of Eqs. (\ref{phqcm1})-(\ref{phqcm2})  are the starting point for the particle-hole-boson treatment of pairing and quartet correlation in the next section. Before presenting the boson approximation, we need to complete the particle-hole description by also expressing the pairing Hamiltonian in terms of particle and hole degrees of freedom.\\

We introduce particle ($i,j,k,\dots$) and hole indices ($a,b,c,\dots$). For hole the subspace, we introduce the pair creation operators as $\tilde{P}^\dagger_a\equiv P_a$. After a decomposition of the pairing Hamiltonian of Eq. (\ref{pairingH}) into particle and hole components we obtain

\bea
\label{pairingHph}
H&=&\sum_{a=1}^{n_p} (2\epsilon_a+V_{aa})\\
&+&\sum_{a=1}^{n_p} (-\epsilon_a-V_{aa})\tilde{N}_a+
\sum_{i=n_p+1}^{N_\text{lev}} \epsilon_i N_i
\nn&+&
\sum_{a,b=1}^{n_p}V_{ab}\tilde{P}^\dagger_a \tilde{P}_b +\sum_{i,j=n_p+1}^{N_\text{lev}}V_{ij}P^\dagger_i P_j
\nn&+&
\sum_{a=1}^{n_p}\sum_{j=n_p+1}^{N_\text{lev}}V_{ai}\left(\tilde{P}_a P_i + P^\dagger_i \tilde{P}^\dagger_a \right)~.
\eea
expressed also in terms of the number of holes operator $\tilde{N}_{a}=2-N_{a}$ which satisfies $\cm{\tilde{N}_a}{\tilde{P}^\dagger_b}=2\delta_{ab}\tilde{P}^\dagger_b$.

For the case of isovector pairing, we analogously introduce the pair creation operators for holes for a given isospin projection as $\tilde{P}^\dagger_{\tau,a}\equiv P_{\tau,a}$ and the corresponding hole number operator $\tilde{N}_{0,a}=4-N_{0,a}$.
Thus, the Hamiltonian (\ref{isovectorH}) decomposes as follows
\bea
H&=&\sum_{a=1}^{q} (4\epsilon_a+3V_{aa})\nn
&+&\sum_{a=1}^{q} (-\epsilon_a-\frac{3}{2}V_{aa})\tilde{N}_{0,a}+
\sum_{i=q+1}^{N_\text{lev}} \epsilon_i N_{0,i}
\nn&+&
\sum_{a,b=1}^{q}V_{ab}\sum_{\tau=\pm1,0}\tilde{P}^\dagger_{\tau,a} \tilde{P}_{\tau,b}
\nn&+&
\sum_{i,j=q+1}^{N_\text{lev}}V_{ij}\sum_{\tau=\pm1,0}P^\dagger_{\tau,i} P_{\tau,j}
\nn&+&
\sum_{a=1}^{q}\sum_{j=q+1}^{N_\text{lev}}V_{ai}\sum_{\tau=\pm1,0}\left(\tilde{P}_{\tau,a} P_{\tau,i} + P^\dagger_{\tau,i} \tilde{P}^\dagger_{\tau,a} \right)~.
\eea

\subsection{Particle-hole boson approximation}

\begin{figure} [ht]
\begin{center}
\includegraphics[width=\columnwidth]{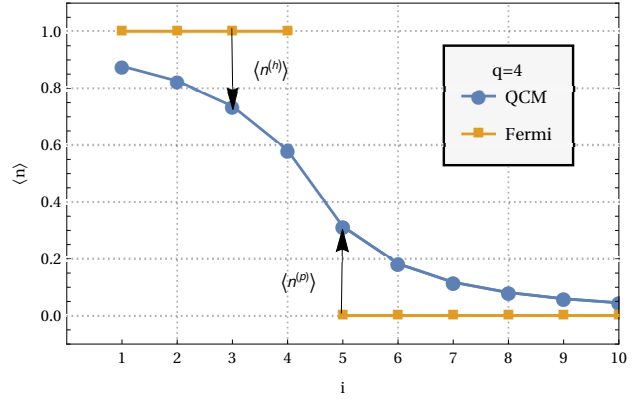}
\caption{The effect of the particle and hole degrees of freedom on the average level occupation fraction $\langle n \rangle$, shown for the QCM in the case of the picket fence model, for $q=4$ quartets.}
\label{fig2}
\end{center}
\end{figure}

The basic idea of our bosonic approximation is to take as reference state the Hartree-Fock state and to describe  the particle and hole degrees of freedom with a simplified bosonic treatment, the particle-hole bosons thus accounting for the deviation from the Fermi distribution (see Fig. (\ref{fig2})). \\

In the following we will first define the particle-hole-boson approximation for the standard pairing case, the generalization to the quarteting case being trivial. 
Our approximation is defined by the replacements for the pair operators and for the vacuum state 
\beq
\begin{aligned}
P^{\dagger}_i &\rightarrow p^\dagger_i~,~\tilde{P}^\dagger_a \rightarrow h^\dagger_a~,~ \fermi \rightarrow \bvac~.
\end{aligned}
\eeq
where the particle and hole bosons annihilate the boson vacuum: $p_i\bvac=0$ and $h_a\bvac=0$, together with the mapping of the particle and hole number operators $\tilde{N}_a\rightarrow\mathcal{N}_{a}$, $N_i\rightarrow\mathcal{N}_{i}$, which also annihilate the bosonic vacuum state, i.e.  $\mathcal{N}_{i}\bvac=0$ and $\mathcal{N}_{a}\bvac=0$. They obey a bosonic algebra
\beq
\label{bosoncomm}
\begin{aligned}
\cm{p_i}{p^\dagger_j}&=\delta_{ij}\pi_j, \cm{h_a}{h^\dagger_b}=\delta_{ab}\eta_b,\cm{p_i}{h^\dagger_j}=0~,\\
\cm{\mathcal{N}_{i}}{p^\dagger_j}&=2\delta_{ij}p^\dagger_j,
\cm{\mathcal{N}_{a}}{h^\dagger_b}=2\delta_{ab}h^\dagger_b~
\end{aligned}
\eeq
where the coefficients $\pi_i$ and $\eta_j$ are c-numbers and all other commutators vanish. Ours is not precisely a traditional boson mapping (in the sense of e.g. Ref. \cite{Gam06} for the same picket fence pairing scenario), but more of an approximate treatment of the fully fermionic model, as the pair operators are now considered to be structureless. Whereas in the original fermionic case, due to the fact that the pairs are composite objects the  pair commutation relation reads $\cm{P_i}{P^\dagger_i}=1-N_i$, where $N_i$ is the number of particles operator,  in the bosonic approximation the pair commutator is just a c-number. Below, we shall distinguish between the pure bosonic case where the commutator is chosen to be unity, and the renormalized case, in which a convenient choice is made in order to account for the effects of the Pauli  exclusion principle.

We then define the corresponding collective bosons
\beq
\label{collectivebosons}
\begin{aligned}
\mathcal{H}^\dagger(y)\equiv\sum_{a=1}^{n_p} y_a h^\dagger_a~,~
\mathcal{P}^\dagger(x)\equiv\sum_{i=n_p+1}^{N_\text{lev}} x_i p^\dagger_i~.\\
\end{aligned}
\eeq
The main point is that we consider the bosonic ground state to be of the same form as the fermionic PBCS condensate
\beq
\label{bPBCS}
\ket{\psi(x,y)}\equiv \sqrt{\chi} \sum_{n=0}^{n_p} \frac{1}{(n!)^2}\left(\mathcal{P}^\dagger(x)~\mathcal{H}^{\dagger}(y)\right)^n\bvac~,
\eeq
where $\chi$ is a normalization constant.
The hole amplitudes $y$ will be compared in the end to the inverse of the fermionic amplitudes corresponding to levels below the Fermi level.
Let us first notice that the commutation relations involving the number of bosons in Eq.(\ref{bosoncomm}) can also be realized 
by using the following replacements
\beq
\begin{aligned}
\mathcal{N}_i\rightarrow \frac{2}{\pi_i} p^\dagger_i p_i~,~~~
\mathcal{N}_a\rightarrow \frac{2}{\eta_a} h^\dagger_a h_a~.\\
\end{aligned}
\eeq
The boson Hamiltonian can be obtained from Eq. (\ref{pairingHph}) as follows
\beq
\label{bosonicH}
\begin{aligned}
H_b&=\sum_{a,b=1}^{n_p}\left(\frac{2\tilde{\epsilon}_a}{\eta_a}\delta_{ab}+V_{ab}\right)h^\dagger_b h_a
\\&+\sum_{i,j=n_p+1}^{N_\text{lev}}\left(\frac{2{\epsilon}_i}{\pi_i}\delta_{ij}+ V_{ij}\right)p^\dagger_i p_j
\\&+\sum_{a=1}^{n_p}\sum_{j=n_p+1}^{N_\text{lev}}V_{ai}\left(h_a p_i + p^\dagger_i h^\dagger_a \right)+\sum_{a=1}^{n_p} (2\epsilon_a+V_{aa})~,
\end{aligned}
\eeq
Throughout this paper, we consistently define the single particle energy corresponding to holes degrees of freedom to be simply $\tilde{\epsilon}_a=-\epsilon_a$, as we neglect the respective interaction contribution appearing in the fully fermionic approach. 
In order to compute the averages of the boson operators on the state (\ref{bPBCS}) it is very convenient to define first the norms of the collective boson pairs
\beq
\label{commcollbospair}
\begin{aligned}
\cm{\mathcal{P}(x)}{\mathcal{P}^\dagger(x)}&=\sum_{i=n_p+1}^{N_\text{lev}}x_i^2\pi_i \equiv S_p~,\\
\cm{\mathcal{H}(y)}{\mathcal{H}^\dagger(y)}&=\sum_{a=1}^{n_p}y_a^2\eta_a \equiv S_h~.\\
\end{aligned}
\eeq
The product $S_pS_h$ will appear frequently in the following and we choose to denote it by $S_{ph}=S_p\cdot S_h$. 
The bosonic approximation is simple enought to allow for an analytical derivation for the norm of the bosonic pair condensate
\beq
\begin{aligned}
\langle \psi(x,y)| \psi(x,y)\rangle 
&=\chi \sum_{n=0}^{n_p} \frac{\left(S_{ph}\right)^n}{(n!)^2} \equiv \chi~ \nu(S_{ph})~.
\end{aligned}
\eeq
The averages of bosonic pair bilinears are easily found to be
\beq
\label{avgbilinear}
\begin{aligned}
\avg{h^\dagger_a h_b}
&=\chi~ y_a \eta_a y_b\eta_bS_p\sum_{n=1}^{n_p}\frac{n}{(n!)^2} S_{ph}^{n-1}~,\\
\avg{p^\dagger_i p_j}&=\chi~ x_i \pi_i x_j\pi_jS_h\sum_{n=1}^{n_p}\frac{n}{(n!)^2} S_{ph}^{n-1} ~,\\
\avg{p_i h_a}&=\avg{p^\dagger_i h^\dagger_a}
={\chi}~x_i\pi_i y_a \eta_a \sum_{n=0}^{n_p-1}\frac{1}{(n!)^2}(S_{ph})^n~.
\end{aligned}
\eeq

Finally, the average of the Hamiltonian over the bosonic pair condensate may thus be written as follows

\beq
\label{universalH}
\begin{aligned}
\avg{H_b}&=\left(\mathcal{H}_{hh} S_p+\mathcal{H}_{pp} S_h\right)\cdot f_1(S_{ph})+\mathcal{H}_{ph} \cdot f_2(S_{ph})\\
&+E_0\cdot \nu(S_{ph}) ~,\\
\mathcal{H}_{hh}&=\sum_{a=1}^{n_p}{2\tilde{\epsilon}_a}{\eta_a}y_a^2 + \sum_{a,b=1}^{n_p}V_{ab} y_a\eta_a y_b \eta_b~,\\
\mathcal{H}_{pp}&=\sum_{i=n_p+1}^{N_\text{lev}}2\epsilon_i {\pi_i}x_i^2 + \sum_{i,j=n_p+1}^{N_\text{lev}}V_{ij} x_i\pi_i x_j \pi_j~,\\
\mathcal{H}_{ph}&=2\sum_{a=1}^{n_p}\sum_{j=n_p+1}^{N_\text{lev}}V_{ai}x_i\pi_i y_a \eta_a ~,
\end{aligned}
\eeq

in terms of the form factors
\beq
f_1(z)=\sum_{n=1}^{n_p}\frac{nz^{n-1}}{(n!)^2} ~,~ f_2(z)=\sum_{n=0}^{n_p-1}\frac{z^{n}}{(n!)^2}~,
\eeq
and the zero point energy $E_0=\sum_{a=1}^{n_p} (2\epsilon_a+V_{aa})$.
It is important to remark that the bosonic approximation remains a highly  nonlinear problem, the particle and hole bosons being coupled not only throught the interaction terms $V_{ai}$, but also through the form factors. We may thus speak of dressed particle and holes degrees of freedom.

The ground state energy corresponding to the minimum of the energy function 
\beq
E(x,y)\equiv \dfrac{\langle\psi(x,y)|{H_b}\ket{\psi(x,y)}}{\langle\psi(x,y)\ket{\psi(x,y)}}~,
\eeq
may be computed upon a minimization procedure with respect to the particle and hole amplitudes $x_i$ and $y_a$.
We note that the energy function has a scaling symmetry $ 
E(x,y)=E\left(\lambda x,\dfrac{y}{\lambda}\right)$, such that the number of independent parameters are actually $N_{lev}-1$.
We will analyze two choices for the commutator coefficients in Eqs. (\ref{bosoncomm})

\begin{enumerate}
\item pure bosonic case: $\eta_a=1$, $\pi_i=1$.

\item renormalized  bosonic case:
\beq
\label{renormcomm}
\begin{aligned}
\eta_a&=1-\frac{1}{2}\avg{\mathcal{N}_a}=1-y_a^2 \eta_aS_p f_1(S_{ph})/\nu(S_{ph})\\
\pi_i&=1-\frac{1}{2}\avg{\mathcal{N}_i}=1-x_i^2 \pi_iS_h f_1(S_{ph})/\nu(S_{ph})
\end{aligned}
\eeq

\end{enumerate}

It follows that in this latter case the commutator coefficients satisfy the following self-consistency condition (as the $S_p$ and $S_h$ terms depend implicitely on them)
\beq
\label{renormcomm2}
\begin{aligned}
\eta_a&=\left(1+y_a^2 S_p f_1(S_{ph})/\nu(S_{ph})\right)^{-1}\\
\pi_i&=\left(1+x_i^2 S_h f_1(S_{ph})/\nu(S_{ph})\right)^{-1}.
\end{aligned}
\eeq
Their precise values may be found, given a set of mixing amplitudes, by a straightforward and rapidly converging iterative procedure.\\

Let us  mention that a renormalized procedure is preferred as to effectively take into account the finite maximum occupation of a given level as dictated by the Pauli exclusion principle. Indeed, it can be seen by combining Eqs. (\ref{renormcomm}) and (\ref{renormcomm2}) that the average level occupation fraction satisfies , e.g. for hole states $\langle n^{(h)}_a\rangle=\dfrac{1}{2}\langle \mathcal{N}_a\rangle=1-\eta_a<1$.\\

The same basic idea of the bosonic approximation for the standard pairing case is easily applicable to the isovector pairing situation. 
Each projection of the triplet of pair operators translates into a corresponding boson
\beq
\begin{aligned}
P^{\dagger}_{\tau,i} &\rightarrow p^\dagger_{\tau,i}~,~ \tilde{P}^\dagger_{\tau, a} \rightarrow h^\dagger_{\tau, a}~,
\end{aligned}
\eeq
where we consider bosonic pairs of different isospin projection to commute:
\beq
\label{bosoncommQ}
\begin{aligned}
\cm{p_{\tau,i}}{p^\dagger_{\sigma,j}}=\delta_{\tau \sigma}\delta_{ij}\pi_j~,~ \cm{h_{\tau,a}}{h^\dagger_{\sigma,b}}=\delta_{\tau \sigma}\delta_{ab}\eta_b
\end{aligned}
\eeq

The expressions of Eqs. (\ref{collectivebosons}) and (\ref{commcollbospair}) are generalized accordingly for each member of the collective pair triplet. 
The bosonic isovector pairing Hamiltonian is basically identical with that of Eq. (\ref{bosonicH}) upon the replacement of the bosonic pair bilinears
with the sum over the three isospin projections, and the redefinition of the constant energy term to $E_0=\sum_{a=1}^q(4\epsilon_a+3V_{aa})$. The bosonic isovector pairing Hamiltonian may thus be cast into the same form as in Eq. (\ref{universalH}). We present in Appendix A the details regarding the form factors in the isovector pairing case.\\

As in the PBCS case, we analyse both choices of pure bosonic commutation relations and their renormalized version. In the latter case, the coefficients are computed for the isovector pairing as 

\beq
\label{etapiQ}
\begin{aligned}
\eta_a&=1-\frac{1}{4}\avg{\mathcal{N}_{0,a}}=1-\frac{1}{2} y_a^2 \eta_aS_p f_1(S_{ph})/\nu(S_{ph})\\
\pi_i&=1-\frac{1}{4}\avg{\mathcal{N}_{0,i}}=1-\frac{1}{2}x_i^2 \pi_iS_h f_1(S_{ph})/\nu(S_{ph})
\end{aligned}
\eeq

We have thus succeeded in applying the same basic idea of approximating as bosons the pairs in the particle-hole expansion of the pair and quartet condensates. We have obtained in both cases the same form of the average of the Hamiltonian on the bosonic version of the condensate, the only differences appearing in the so-called form factors.

\begin{widetext}
\section{Numerical application}

\begin{figure} [ht]
\begin{center}
\includegraphics[width=\columnwidth]{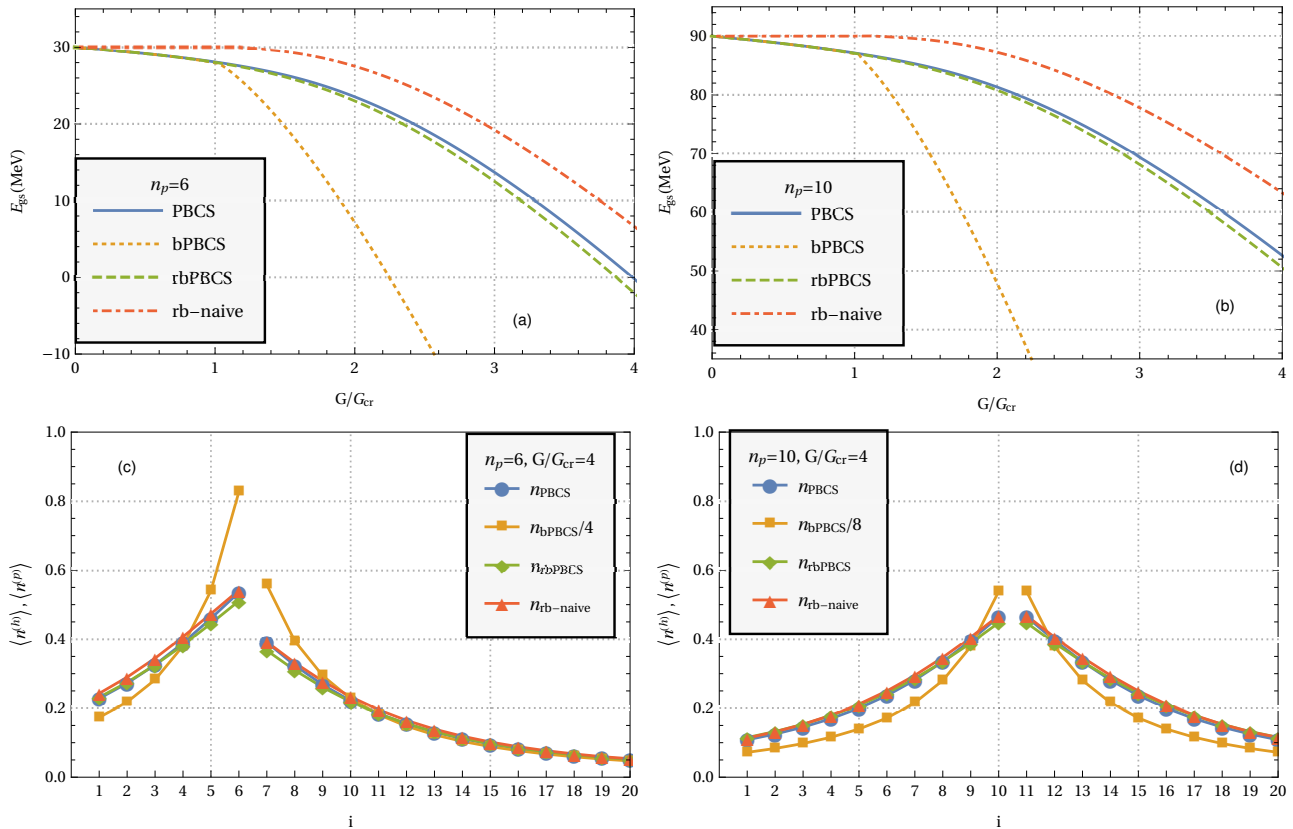}
\caption{The ground state energies in MeV vs the  ratio $G/G_\text{cr}$  for $n_p=6$ (a) and $n_p=10$ (b) and the average level occupation fractions vs state index  for $n_p=6$ (c) and $n_p=10$ (d) for the PBCS, bPBCS and rbPBCS approaches of the picket fence model.}
\label{fig3}
\end{center}
\end{figure}

\begin{figure} [ht]
\begin{center}
\includegraphics[width=\columnwidth]{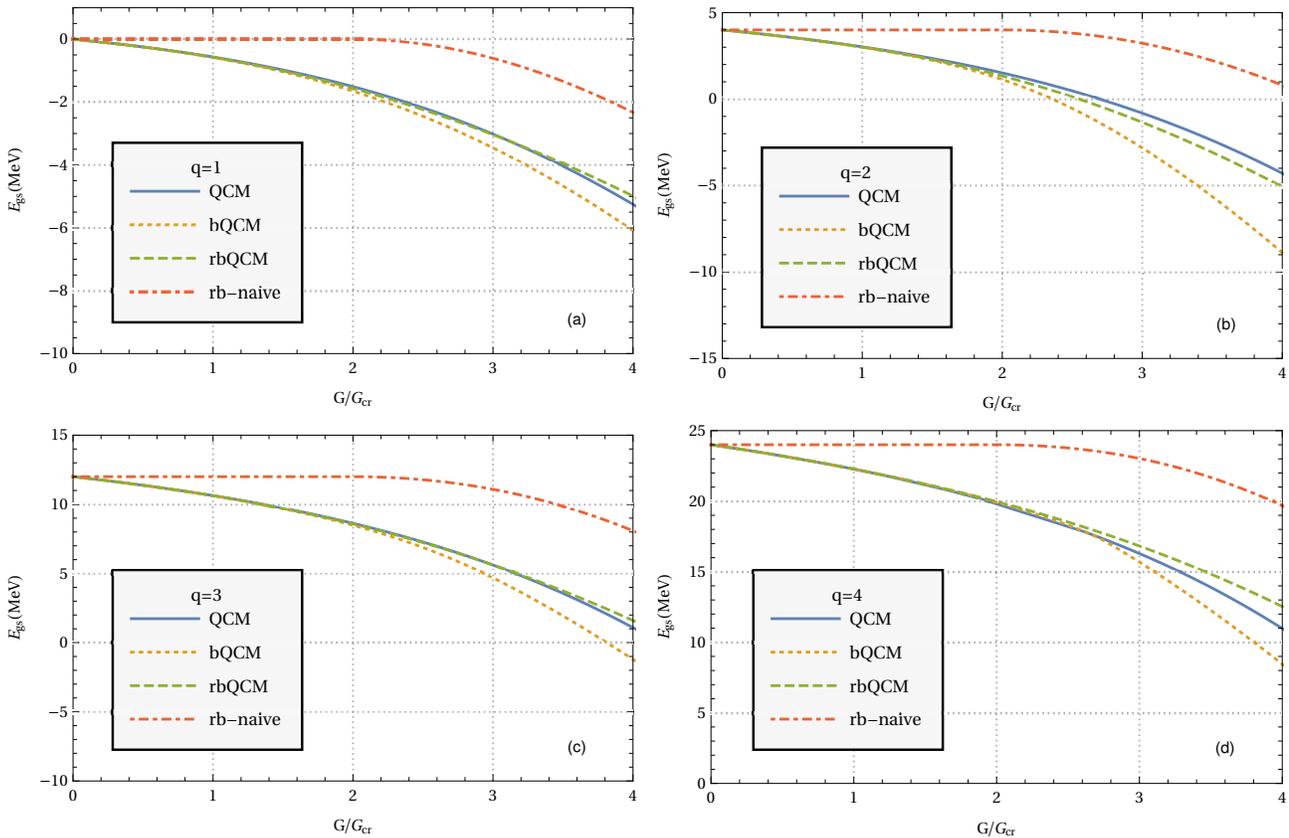}
\caption{The ground state energies (in MeV) vs the ratio $G/G_\text{cr}$ for $q=1$ (a), $q=2$ (b), $q=3$ (c), $q=4$ (d) the QCM, bQCM and rbQCM approaches of the picket fence model.}
\label{fig4}
\end{center}
\end{figure}

\begin{figure} [ht]
\begin{center}
\includegraphics[width=\columnwidth]{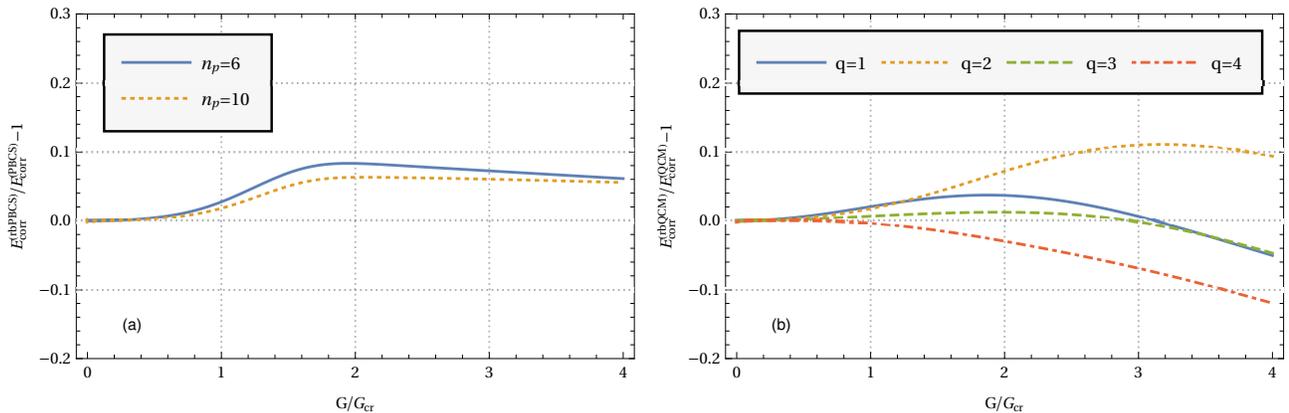}
\caption{The error in the correlation energy of the renormalized boson approximation rbPBCS (a) and, respectively, rbQCM (b), relative to the fermionic cases of PBCS (a) and QCM (b).}
\label{fig5}
\end{center}
\end{figure}

\begin{figure} [ht]
\begin{center}
\includegraphics[width=\columnwidth]{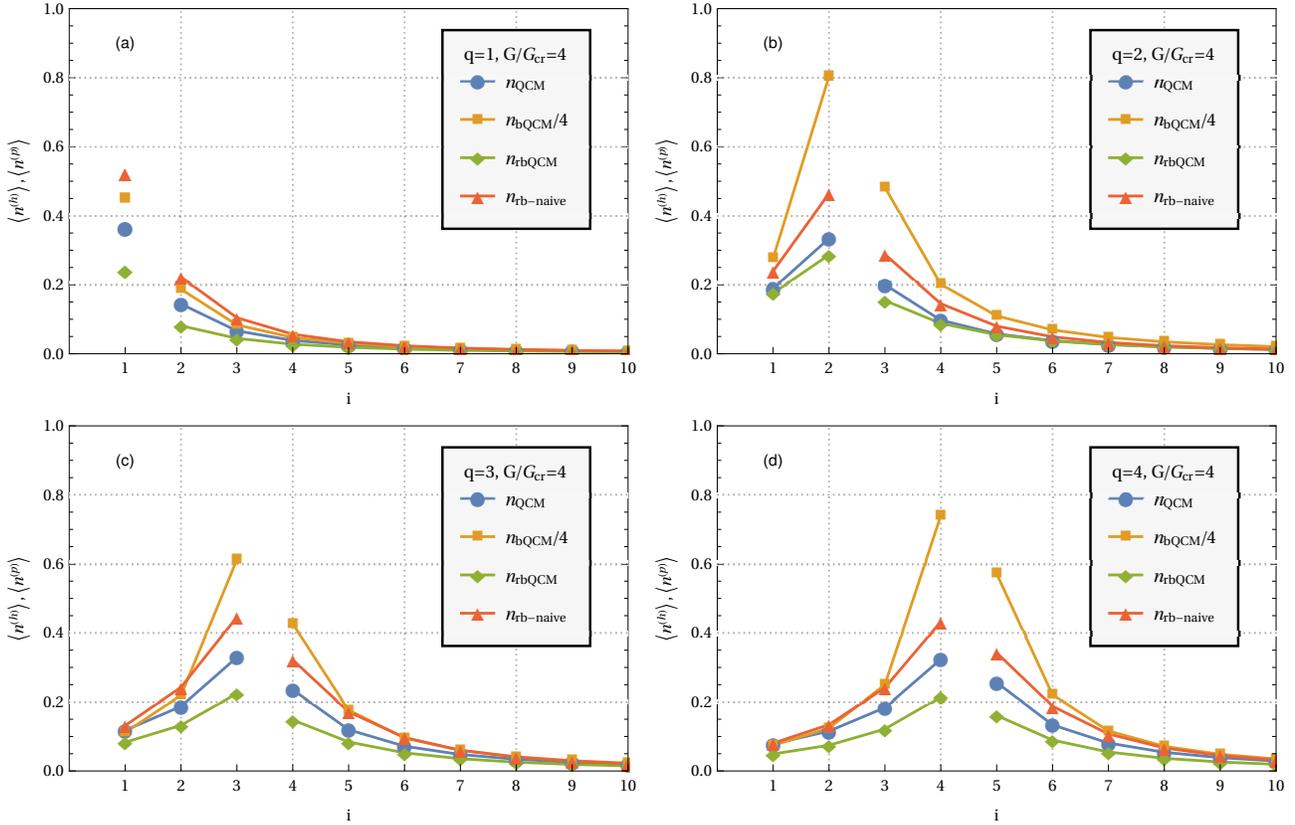}
\caption{Average level occupation fractions vs state index for the QCM, bQCM and rbQCM for the first 10 levels of the picket fence model, for $q=1$ (a), $q=2$ (b), $q=3$ (c) and $q=4$ (d) in the strong pairing regime at $G/G_\text{cr}=4$.}
\label{fig6}
\end{center}
\end{figure}

\begin{figure} [ht]
\begin{center}
\includegraphics[width=\columnwidth]{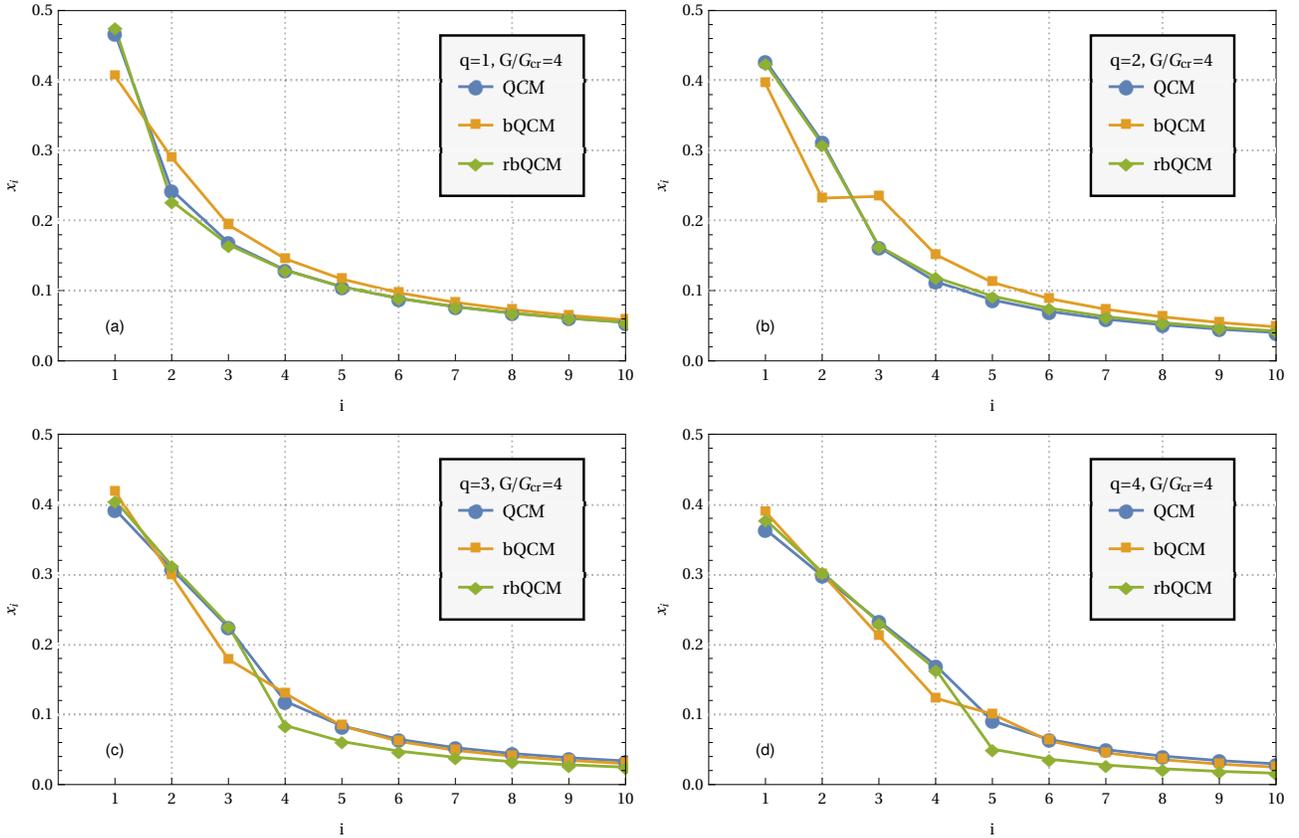}
\caption{Mixing amplitudes $x_i$ vs state index for the QCM, bQCM and rbQCM for the first 10 levels of the picket fence model,  for $q=1$ (a), $q=2$ (b), $q=3$ (c) and $q=4$ (d) in the strong pairing regime at $G/G_\text{cr}=4$. The best fit for the amplitudes in the bosonic models was chosen as allowed by the scaling symmetry.}
\label{fig7}
\end{center}
\end{figure}

\begin{figure} [ht]
\begin{center}
\includegraphics[width=\columnwidth]{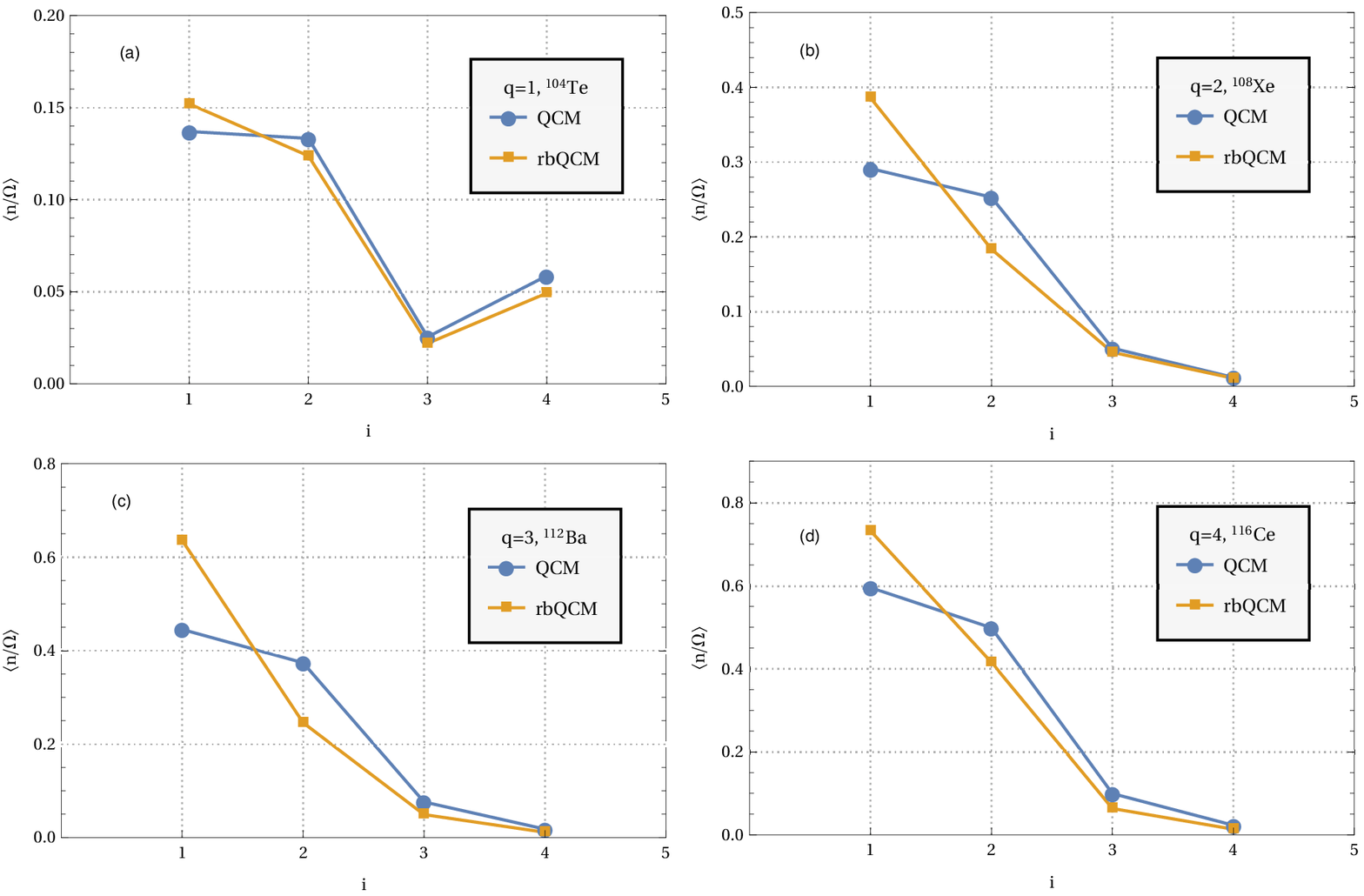}
\caption{ Average level occupation fractions $\langle n_i/\Omega_i\rangle\equiv\langle N_{0,i}\rangle/(2(2j_i+1))$ versus the state index $i$ corresponding to the exact QCM and to the renormalized bosonic approximation rbQCM for the nuclei above the  $^{100}$Sn core, computed with the Bonn A effective interaction of Ref. \cite{Jen95}. }
\label{fig8}
\end{center}
\end{figure}

\end{widetext}

We have analyzed the projected-BCS cases of $n_p=6$  and $n_p=10$ pairs and the QCM cases of $q=1,2,3,4$ quartets distributed over 20 equally spaced, $\epsilon_k=(k-1)$ MeV, doubly-degenerate single particle levels, interacting via a constant pairing force, $V_{ij}=-G$. The solutions for the QCM fermionic approach were obtained by using the analytical method described in \cite{qcmanalitic}. For the PBCS case, we implemented the recurrence relations presented in \cite{Duk00}. In all cases the minimization of the energy function with respect to the mixing amplitudes was carried out by using the e04ucf routine of the NAG library. \\

In the following, we shall denote the pure bosonic approximation, corresponding to $\eta_a=1, \pi_i=1$, by "bPBCS" in the standard pairing case and by "bQCM" for the isovector pairing case. The renormalized versions, defined by the prescriptions of Eq. (\ref{renormcomm2}) and Eq. (\ref{etapiQ}), are referred to as "rbPBCS" and, respectively, "rbQCM".\\

The pairing strength $G$, for both standard pairing and isovector pairing cases, is given in units of the critical strength $G_\text{cr}$ for which the pairing gap vanishes in the standard BCS and, respectively,  proton-neutron BCS \cite{Cam03, Che67, Del10a}. This allows to distinguish between the weak, medium, and strong pairing regimes which roughly correspond to $G<G_\text{cr}$, $G\sim G_\text{cr}$ and $G>G_\text{cr}$. For our particular picket-fence model, the specific values are $G_\text{cr}^{(n_p=6)}=0.238$ MeV, $G_\text{cr}^{(n_p=10)}=0.234$ MeV, and respectively $G_\text{cr}^{(q=1)}=0.144$ MeV, $G_\text{cr}^{(q=2)}=0.132$ MeV, $G_\text{cr}^{(q=3)}=0.127$ MeV, $G_\text{cr}^{(q=4)}=0.123$ MeV. \\

In all cases, we find excellent agreement between the fermionic and both bosonic approximations in the weak pairing regime. However, as the strength of the correlations increases the inadequacy of the pure bosonic approach is quickly revealed, the main reason being its inabilty to reproduce the finite level occupancy as dictated by the Pauli principle. Indeed, we observe from Fig. \ref{fig3} (b) and (c) that the average level occupation fractions for the pure bosonic approach in the PBCS case, for a strong pairing scenario, exceed unity for the states close to the Fermi surface. As a consequence, the ground state energy is not correctly reproduced. These problems are however completely solved by the renormalization procedure described in the previous section. Indeed, the rbPBCS approach is in almost perfect agreement with the fully fermionic results regarding average level occupation fractions, as shown in the same figure. The renormalization restrictions also have the effect of bringing the ground state energy much closer to the fermionic value, as displayed in  Fig. \ref{fig3} (a) and (b).\\

In order to emphasize the essential role of considering the bosonic ground state as having the same structure as the particle-hole version of the fermionic condensate (see Eqs. (\ref{PBCS}) and (\ref{bPBCS})), in Fig. \ref{fig3}  (a) and (b) we also plotted the ground state energy in the so-called naive renormalized bosonic approach (denoted in the following as "rb-naive). In this case, we applied the bosonization procedure (in the renormalized version) directly to the original PBCS condensate of Eq. (\ref{PBCS0}) and to the original pairing Hamiltonian of Eq. (\ref{pairingH}). Regarding the ground state energy, we notice strong discrepancies between the naive bosonic approach and the fermionic case, for all interaction strengths (except the $G=0$ case which is reproduced due to the renormalization procedure). In this way we support the idea that the particle-hole expansion of the (pair) condensate contains a significant amount of information about the pure fermionic correlations in the ground state. Let us mention that the adequacy of the particle hole description of pairing correlations is indicated by the fact that the basic physical behavior is determined by the fluctuations around the Fermi state, small in the weak pairing regime and larger in the medium and strong regimes.  It is however remarkable that even in the naive approach the average level occupancies are reproduced to a high degree of accuracy with respect to the fermionic case, which is an indication of the effectiveness of the renormalization prescription in accounting for the effects of the Pauli exclusion principle.\\

The situation is qualitatively similar in the case of quarteting correlations, as seen in Fig. \ref{fig4}.
 There is a very good agreement between the fermionic and bosonic approaches in the weak and medium pairing regimes. In the strong pairing case however, the correlation energy is  overestimated in the pure bosonic approach (and even more so in the naive bosonic version), and slightly underestimated within the renormalized bosonic rbQCM approximation. Let us note that the numerical value of the critical strenght in the isovector pairing case is about half of the value corresponding to the single species pairing case. This translates into fact that the pairing strength relative to the level spacing, $G/\Delta\epsilon$ which may also be used as an alternative indicator of the intensity of pairing correlations, differs by a factor of two in the isovector and standard pairing cases. Indeed, the ground state energy shows more variation in the standard pairing case than in the isovector pairing case, up to the same value of $G/G_{cr}=4$. The errors in the bosonic approaches increase beyond this point, in a similar way to the behavior presented in Fig. (\ref{fig3}.a) and (\ref{fig3}.b).  \\
 
 We show in Fig. \ref{fig5} the errors in the correlation energy of the renormalized bosonic approximation relative to the fermionic approaches, i.e. for PBCS in the left panel (a) and respectively for QCM in the right panel (b). We note the perfect agreement in the weak pairing regime in all cases. Slight discrepancies start to emerge in the medium pairing regime, but all errors are at most of the order of 10\% even in the strong pairing scenario, up to $G/G_\text{cr}=4$.\\
 
  Regarding the average level occupation fraction in the quarteting case, we notice in Fig. \ref{fig6}, as
expected, unphysical values that exceed unity for the pure bosonic approach. On the other hand, the rbQCM approximation results sligthly underestimate the exact values. The more pronounced deviations of the isovector pairing bosonic approximations with respect to the QCM fermionic approach are to be traced back to the fact that in these cases we made the additional assumption of commuting pairs of different isospin projection (see Eq. (\ref{bosoncommQ})). While it is not difficult to conceive further improvements that take into account the pair mixing effects within a bosonic treatment, we limit ourselves in this work is to the assessment of the consequences of the simplest kind of approximation.  We also remark that an opposite behavior is exhibited by the naive renormalized bosonic approach in the quarteting case, i.e. the results are slightly overestimated. We expect that the results will be comparable to those obtained in the PBCS scenario once pair mixing effects are taken into account.\\

We then compare the results regarding the mixing amplitudes for the quarteting case in the strong pairing regime. One should recall that in the bosonic case they are defined up to an overall factor, due to the scaling symmetry of the bosonic ground state energy mentioned in the previous seection, and to the lack of the unit norm constraint in the bosonic approach (the normalization constant is this case beint a free parameter). We thus limit ourselves to a comparison of the relative behavior of the fermionic and bosonic amplitudes, by choosing the overall factor in the bosonic case as to give the best fit with respect to the fermionic case. In the following, for the purpose of this comparison, we need to work with the inverse amplitudes for the hole states (see also the discussion following Eq. (\ref{bPBCS})), thus we perform the replacement $x_a^{(b)}\rightarrow 1/x_a^{(b)}$ for the bosonic hole amplitudes. Explicitely, a least squares fit for $x^{(f)}=\alpha x^{(b)}$ leads to the the expression $\alpha=\left[\sum_i x_i^{(f)} x_i^{(b)}\right]/\left[\sum_i \left(x_i^{(b)}\right)^2\right]$, where $f$ stands for the fermionic QCM case and $b$ refers to each of the two bosonic approximations. \\

As seen from Fig. \ref{fig7}, a very good agreement regarding the behavior of the mixing amplitudes is found between the exact QCM result and the renormalized bosonic rbQCM approximation for $q=1$ and $q=2$, even in the strong pairing regime; the pure bosonic theory, however, shows discrepancies, especially for the states around the Fermi level. For a larger number of quartets, the results corresponding to the renormalized bosonic version are not better than the ones for the pure bosonic case. This is caused by the diagonal approximation of the pair mixing matrix in isospin space, Eq. (\ref{bosoncommQ})), which for the renormalized version generally leads to  a slight underestimation of the results for particle states, and a slight overestimation for hole states, regarding both mixing amplitudes and average  level occupancies (of physical particles). \\

Finally, we present, as a more realistic application of our approximation, the computation of the ground state correlations in the nuclei above $^{100}$Sn, namely $^{104}$Te, $^{108}$Xe, $^{112}$Ba and $^{116}$Ce, corresponding to a number of quartets $q=1,2,3,4$ in the $sdg$ shell. We consider the same model space and interaction as in Ref. \cite{San12}, namely the spherical spectrum $\epsilon_{2d_{5/2}}=0.0$MeV,  $\epsilon_{1g_{7/2}}=0.2$MeV,  $\epsilon_{2d_{3/2}}=1.5$MeV and  $\epsilon_{3s_{1/2}}=2.8$MeV together with the effective Bonn A isovector pairing potential of Ref. \cite{Jen95}. The results regarding the correlations energy are summarized in Table I. In the exact fermionic approach, we reproduce, within numerical accuracy, the results of Ref. \cite{San12}. In the renormalized bosonic approximation the results are good, showing a relative error of the order of 10\% with respect to the fermionic case. The comparison of the two approaches regarding the average level occupation fraction is presented in Fig. (\ref{fig8}), where we plot the occupancy $\langle N_{0,i}\rangle/(2(2j_i+1))$ versus the spherical state index $i$. We find a very good agreement for $q=1$, however for a larger number of quartets we notice, for the boson approach, the usual underestimation of particle and hole occupancies (which translates to an overestimation of physical particle occupancies for states below the Fermi level). A more detailed analysis of the boson approximation as applied to realistic pairing scenarios will be perfomed in future works, in the context of a more accurate treatment of bosonic pair mixing in isospin space.\\

 Let us also remark upon the interesting possibility of extending our approach to a more general two-body Hamiltonian: the starting point of our discussion was that the trial ground state had a simple condensate type expression which could be nicely reformulated as a particle-hole expansion. Unfortunately, this is not true for the ground state of a general Hamiltonian. However, we have noticed the suitability of the renormalization procedure for the boson degrees of freedom, which accounts very well for the effects of the exclusion principle. This fact is very promising, as the various contributions arising in the standard bosonic expansions can be effectively resummed within the renormalization procedure, thus keeping the boson mapping to its simplest form. The actual degree to which this equivalence is precisely realized is a very interesting question which will be explored in future works.  \\

\begin{table}
\label{table1}
\caption {Correlation energies corresponding to the exact QCM and to the renormalized bosonic approximation rbQCM (together with the relative errors in the round brackets), for the nuclei above the  $^{100}$Sn core, computed with the Bonn A effective interaction of Ref. \cite{Jen95}.  } 
\begin{tabular}{lll}
\hline\hline
         & QCM  &~~~ rbQCM   \\    \hline
	$^{104}$Te&      3.847  & 3.445 (10.4\%)\\
	 $^{108}$Xe&   6.726 & 6.512 (3.2\%)\\
	   $^{112}$Ba&   8.629 & 7.470 (13.4\%) \\
	   $^{116}$Ce&    10.332 & 9.225 (10.7\%)\\ \hline\hline         

\end{tabular}

\end{table}

It is noteworthy that in the isovector pairing case, the QCM offers an almost perfect description of the ground state (the correlation energies  are generally within a 1\% error with respect to the exact shell model diagonalization \cite{San12}), even for the weak pairing regime (as opposed to the PBCS case (see e.g. \cite{Duk16}). As such, the renormalized bosonic treatment of the quartet condensates promises to be a simple but also quite a precise approach.\\

\section{Conclusions}

We developed a new bosonic approximation for pair and quartet condensates, corresponding to the standard pairing and isovector pairing scenarios. \\

The starting point was the reformulation of the pair and quartet condensates as a particle hole expasion with respect to the Hartree-Fock state.
In  particular, we derived the expression of the quartet condensate state as a particle-hole expansion, and found both quartet-quartet excitations and coupled pair excitations. We evidenced the remarkable fact that for both standard pairing and for the more complicated quarteting correlations there is an inverse $x$ versus $1/x$ symmetry for particle and hole mixing amplitudes, which is a strong argument for the existence of a quasiparticle representation for quartet systems.\\

We then introduced a straightforward bosonic formalism which is very similar in both pairing and quarteting cases. The average of the Hamiltonian on the condensate states has the same form in both cases, the only differences being in the expressions of the above defined form-factors. 
We have studied both the pure bosonic approach 
and the renormalized version, and we have compared the particle-hole bosonic version to the "naive" prescription where we applied the boson approximation directly to the original condensate state, without performing the particle-hole reformulation. We have found a good 
agreement between the fermionic and the renormalized particle-hole bosonic approach in the case of a picket fence model of doubly degenerate states and in a realistic shell model space with the Bonn A effective isovector pairing potential for the nuclei above the $^{100}$Sn core. We note however that in the quarteting case the pair mixing effects have been neglected for simplicity. Their contribution is expected to increase the accuracy of the boson approximation, as it will be shown in future works.  \\

In conclusion, we have found that the particle-hole expansion of the pair and quartet condensates contains a lot of information about the fermionic correlations in the ground state, which allows for a good description in terms of bosonic degrees of freedom (provided one effectively takes into account the exclusion principle via the renormalization procedure).

\begin{widetext}
\appendix
\section{Particle-hole expansion of the PBCS and QCM condensate states}

In order to derive the particle hole formulation of the pair condensate, we first express the Hartree-Fock state of Eq. (\ref{fermi1}) in terms of the hole component of the coherent pair as
\beq
\fermi=\frac{1}{n_p!}\frac{1}{\Pi_1} \left(\Gamma^{\dagger}_h(x)\right)^{n_p}\ket{0}~,
\eeq
where $\Pi_1=x_1x_2\cdots x_{n_p}$. The main trick is then to use a coherent pair of inverse arguments $\Gamma_h\left(\dfrac{1}{x}\right)$. 
Starting from the commutator $\cm{P_{i}}{P^\dagger_j}=\delta_{ij}(1-\hat{N}_i)$, it is easy to compute
\bea
\cm{\Gamma_h\left(\dfrac{1}{x}\right)}{\Gamma^\dagger_h(x)}&=
&n_p-\sum_{i=1}^{n_p}\hat{N}_i\equiv n_p -\hat{N}_h~.
\eea
From this it may be shown that
\beq
\left(\Gamma_h\left(\dfrac{1}{x}\right)\right)^j\left(\Gamma^\dagger_h(x)\right)^k\ket{0}=\frac{j! k!}{(k-j)!}\left(\Gamma^\dagger_h(x)\right)^{k-j}\ket{0}~.
\eeq
For the particular case of $k=n_p$, we may relate the action of the coherent pair of inverse arguments on the Hartree-Fock state to the action of the original coherent pair on the $\ket{0}$ vacuum as
\beq
\left(\Gamma_h\left(\dfrac{1}{x}\right)\right)^j\fermi=\frac{1}{\Pi}\frac{j!}{(n_p-j)!}\left(\Gamma^\dagger_h(x)\right)^{n_p-j}\ket{0}~.
\eeq

By employing this expression in the expansion of the PBCS condensate, we arrive at  the form mentioned in Eq. (\ref{PBCS})
\beq
\begin{aligned}
\ket{PBCS}&=\left(\Gamma^\dagger_h(x)+\Gamma^\dagger_p(x)\right)^{n_p}\ket{0}
=n_p!\cdot \Pi_1\cdot  \sum_{j=0}^{n_p} \frac{1}{(j!)^2}\left(\Gamma^\dagger_p(x)~\Gamma_h\left(\dfrac{1}{x}\right)\right)^j\fermi~,
\end{aligned}
\eeq

For the quarteting case, we perform a similar maneuver. We start from the the $q$-quartet condensate of Eq. (\ref{quartetexpans}) state which may be expanded as follows

\beq
\label{psiqexpans}
\begin{aligned}
\ket{\Psi_q}=\sum_{n=0}^q\sum_{j=0}^n\frac{q!}{(n-j)! j! (q-n)!}2^j\left(Q^\dagger_p\right)^{n-j} \left[\Gamma^\dagger_{p}\Gamma^\dagger_{ h}\right]^j \left(Q^\dagger_h\right)^{q-n}\ket{0} 
\end{aligned}
\eeq

We will perform the transition to the particle hole representation in two steps:
\begin{enumerate}
\item we first express the $n$-hole-quartet state $\left(Q^\dagger_h\right)^{q-n}\ket{0} $ as the annihilation of $n$ quartets from the Hartree-Fock state of Eq. (\ref{fermi2}) and
\item we then perform a similar computation for the term involving the coupled pairs. Note that, as in the PBCS case, the collective annihilation operators will dependend on the inverse amplitudes.
\end{enumerate}

\subsection{{{\textbf Relate}} $\left(Q^\dagger_h(x)\right)^{q-n}\ket{0}\sim \left(Q_h\left(\dfrac{1}{x}\right)\right)^{n}|{HF}\rangle$}

Using the hole quartet $Q^\dagger_h$ we may also express the Hartree-Fock state as
\beq
\label{fermiQ}
\fermi=\frac{1}{\Pi_2}\frac{2^q}{(2q+1)!}\left(Q^\dagger_h(x)\right)^q\ket{0}~,
\eeq
where $\Pi_2=x_1^2\cdot x_2^2 \cdots  x_q^2$. Consider the action of the collective pair annihilation operators of inverse amplitudes on a n-quartet state on the hole subspace. From the SO(5) algebra \cite{danielphd} of the hole operators, it is not difficult to show that
\beq
\label{commpairquartet}
\Gamma_{\tau,h}\left(\frac{1}{x}\right)Q^\dagger_{h}(x)^n\ket{0}=(-1)^{1-\tau} \cdot n \cdot (2q-2n+3)\cdot \Gamma^\dagger_{-\tau,h}(x)Q^\dagger_{h}(x)^{n-1}\ket{0}~.
\eeq
From this relation it follows that the action of a collective quartet annihilation operator on an $n$-hole-quartet state is
\beq
Q_{h}\left(\frac{1}{x}\right)Q^\dagger_{h}(x)^n\ket{0}= n \cdot (2q-2n+3)\cdot \left[(n-1)(2q-2n+5)-3(q-2n+2)\right]\cdot Q^\dagger_{h}(x)^{n-1}\ket{0}~.
\eeq
By iterating this relation, we obtain
\beq
\left(Q_{h}\left(\frac{1}{x}\right)\right)^a Q^\dagger_{h}(x)^n\ket{0}= \frac{ (2n+1)!\cdot (2a+1)!}{2^{2a}~(2n-2a+1)!}Q^\dagger_{h}(x)^{n-a}\ket{0}~.
\eeq
It is now possible to compute the action of hole-quartet annihilation  operators on the Hartree-Fock state
\beq
\Pi_2\frac{2^a}{(2a+1)!} \left(Q_{h}\left(\frac{1}{x}\right)\right)^a \fermi =\frac{2^k}{(2k+1)!} \left(Q^\dagger_{h}\left({x}\right)\right)^k \ket{0},~~~a+k=q~,
\eeq
which allows for the first rewriting of the $q$-quartet  condensate state as
\beq
\label{psiqexpans2}
\begin{aligned}
\ket{\Psi_q}=\Pi_2\sum_{n=0}^q\sum_{j=0}^n\frac{q!}{(n-j)! j! (q-n)!}2^{2n-q}\frac{(2q-2n+1)!}{(2n+1)!}2^j\left(Q^\dagger_p(x)\right)^{n-j} \left[\Gamma^\dagger_{p}(x)\Gamma^\dagger_{ h}(x)\right]^j \left(Q_h\left(\dfrac{1}{x}\right)\right)^{n}\fermi~.
\end{aligned}
\eeq

\subsection{\textbf{Relate $\left[\Gamma^\dagger_{p}\Gamma^\dagger_{ h}\right]^j \left(Q_h\right)^{n}|{HF}\rangle \sim \left[\Gamma^\dagger_{p}\Gamma_h\right]^j \left(Q_h\right)^{n-j}|{HF}\rangle $}}

In the following we denote the collective annihilation operators on the hole subspace simply by $A_h\equiv A_h\left(\dfrac{1}{x}\right)$. We also use the notation
\beq
\left[ \Gamma^\dagger_p \Gamma_h\right] \equiv \Gamma^\dagger_{1,p}\Gamma_{1, h}+\Gamma^\dagger_{-1,p}\Gamma_{-1, h}+\Gamma^\dagger_{0,p}\Gamma_{0, h}~.
\eeq
Let us compute the commutator
\beq
\cm{\left[ \Gamma^\dagger_p \Gamma_h\right]}{\left[\Gamma^\dagger_{p}\Gamma^\dagger_{ h}\right]}=Q^\dagger_p~(q-\frac{1}{2}\hat{N}_{0,h})~,
\eeq
where $\hat{N}_{0,h}=\sum_{i=1}^{q}\hat{N}_{0,i}$ is the total number of particles on the hole subspace.  From the previous relation it follows that
\beq
\label{commcoupledpairs}
\cm{\left[ \Gamma^\dagger_p \Gamma_h\right]}{\left[\Gamma^\dagger_{p}\Gamma^\dagger_{ h}\right]^n}=Q^\dagger_p~\left[\Gamma^\dagger_{p}\Gamma^\dagger_{ h}\right]^{n-1}\cdot \frac{n}{2}(2q-n+1-\hat{N}_{0,h}).
\eeq
Combining Eqs. (\ref{commpairquartet}) and (\ref{commcoupledpairs}) we obtain the recurrence relation
\beq
\begin{aligned}
\ket{{jn}}&\equiv \left[\Gamma^\dagger_{p}\Gamma^\dagger_{ h}\right]^j \left(Q_h\right)^n \fermi\\
&=n(2q-2n+3)\left(\left[ \Gamma^\dagger_p \Gamma_h\right]\ket{j-1,n-1}+\frac{j-1}{2}(2q+j-4n+2)Q^\dagger_p \ket{j-2,n-1}\right)~.
\end{aligned}
\eeq
A careful analysis of this recurrence relation leads to the expression
\beq
\label{pairq}
\begin{aligned}
\left[\Gamma^\dagger_{p}\Gamma^\dagger_{ h}\right]^j \left(Q_h\right)^n \fermi&=\frac{n!}{(n-j)!}\frac{(2q-2n+2j+1)!}{(2q-2n+1)!}\frac{(q-n)!}{2^j(q-n+j)!}\\
&\sum_{k=0}^{\left[j/2\right]}\frac{1}{2^kk!}\frac{j!}{(j-2k)!}(q-2n+j)_k\left[ \Gamma^\dagger_p \Gamma_h\right]^{j-2k} \left(Q^\dagger_pQ_h\right)^k \left(Q_h\right)^{n-j} \fermi~.
\end{aligned}
\eeq
where the notation $(z)_n$ is the Pochammer symbol $(z)_n=z(z-1)\cdots(z-n+1)$ and the $[n]$ is floor function.\\

By using the expressions (\ref{psiqexpans2}) and (\ref{pairq}) we obtain
\beq
\begin{aligned}
\ket{\Psi_q}=&\Pi_2\sum_{n=0}^q\sum_{j=0}^n\frac{q!}{((n-j)!)^2}2^{2n-q}\frac{n! (2q-2n+1)!}{(2n+1)!(q-n+j)!}
\sum_{k=0}^{\left[j/2\right]}\frac{(q-2n+j)_k}{2^kk!(j-2k)!}
\left(Q^\dagger_p Q_h\right)^{n-j+k} \left[\Gamma^\dagger_{p}\Gamma_{ h}\right]^{j-2k} \fermi~.
\end{aligned}
\eeq
Let us mention that Eqs. (\ref{phqcm1}-\ref{phqcm2}) follow after regrouping the terms of the quartet-quartet and coupled pairs excitations
with the same powers.

\section{Pairing and quarteting bosonic form factors}

In the standard pairing case, the form factors and norm function entering the expression of the Hamiltonian expectation value on the bosonic version of the condensate (see Eq. (\ref{universalH})) may be computed analytically for any number of pairs. They can be read off the averages of bosonic pair bilinears of Eqs. (\ref{avgbilinear}),

\beq
\begin{aligned}
f_1(z)&=\sum_{n=1}^{n_p}\frac{nz^{n-1}}{(n!)^2}~,~f_2(z)=\sum_{n=0}^{n_p-1}\frac{z^{n}}{(n!)^2}~,~\nu(z)= \sum_{n=0}^{n_p} \frac{z^n}{(n!)^2}  ~.
\end{aligned}
\eeq

However, due to the more complicated form of the quartet condesate particle-hole expansion of Eqs. (\ref{phqcm1})-(\ref{phqcm2}) we are unable to provide general analytical expressions of the form factors for the isovector pairing case. For each particular number of quartets, they may be computed in a straightforward fashion by expanding the collective quartets and coupled bosonic pairs into individual pairs and then evaluating each expression, taking advantage of the bosonic character of the pairs. \\

We present below the formulas for the form factors and norm function in the isovector pairing case for a number of quartets ranging from one to four.

\beq
\begin{aligned}
f_1^{(q=1)}(z)&=12 + 8z~,\\
f_2^{(q=1)}(z)&=18 + 8z~,\\
\nu^{(q=1)}(z)&=9 + 12z + 4z^2~.
\end{aligned}
\eeq

\beq
\begin{aligned}
f_1^{(q=2)}(z)&=1200 + 1440z + 576z^2 + 64z^3~,\\
f_2^{(q=2)}(z)&=1800 + 2400z + 1056z^2 + 128z^3~,\\
\nu^{(q=2)}(z)&=900 + 1200z + 720z^2 + 192z^3 + 16z^4~.
\end{aligned}
\eeq

\beq
\begin{aligned}
f_1^{(q=3)}(z)&=529200 + 1734048z + 302400z^2 + \frac{4262976}{49}z^3 + 8640z^4 + 384z^5~,\\
f_2^{(q=3)}(z)&=793800 + 1421280z + 997920z^2 +   58752z^3 - 7776z^4 + 1152z^5~,\\
\nu^{(q=3)}(z)&=396900 + 529200z + 867024z^2 +   100800z^3 + \frac{1065744}{49}z^4 +  1728z^5 + 64z^6~.
\end{aligned}
\eeq

\beq
\begin{aligned}
f_1^{(q=4)}(z)&=2048 z^7+86016 z^6+14678016 z^5+\frac{179532800 z^4}{3}+101606400 z^3+1758827520 z^2+4854753792 z\\
&+685843200~,\\
f_2^{(q=4)}(z)&=8192 z^7-718848 z^6+11515904 z^5-47715840 z^4+539965440 z^3+3803466240 z^2+2347107840 z\\
&+1028764800~,\\
\nu^{(q=4)}(z)&=256 z^8+12288 z^7+2446336 z^6+\frac{35906560 z^5}{3}+25401600 z^4+586275840 z^3+2427376896 z^2\\
&+685843200 z+514382400~.
\end{aligned}
\eeq
\end{widetext}

\acknowledgements

This work was supported by the grants of the Romanian Ministry of Research and Innovation, CNCS - UEFISCDI, PN-
III-P4-ID-PCE-2016-0092, PN-III-P4-ID-PCE-2016-0792, within PNCDI III, and PN-19060101/2019.

\end{document}